\documentstyle[aps]{revtex}
\begin{document}
\title{Local off-cubic distortion \\
the cause for the low- and high-spin states of the Co$^{3+}$ ion* }
\author{R.J. Radwa\'{n}ski}
\address{Center for Solid State Physics, \'{s}w. Filip 5, 31-150 Krak\'{o}w.\\
Inst. of Physics, Pedagogical University, 30-084 Krak\'{o}w. }
\author{Z. Ropka}
\address{Center for Solid State Physics, \'{s}w. Filip 5, 31-150 Krak\'{o}w, Poland.\\
email: sfradwan@cyf-kr.edu.pl}
\maketitle

\begin{abstract}
It is pointed out that it is the local symmetry that determines the
realization of the low- and high-spin state of the Co$^{3+}$ ion. We can
rigorously prove it treating the Co$^{3+}$ ion as the highly-correlated
electron system 3d$^6$, that fulfiles the Hund's rules with taking into
account the spin-orbit coupling. As the function of the sign of the
off-cubic crystal-field distortion the magnetic or non-magnetic state is
realized.



Receipt date by Phys.Rev.Lett. 29.03.1999
\end{abstract}

\date{(29.03.1999)}

The origin of low- and high-spin states of the Co$^{3+}$ ion formed in
different ionic compounds is under the long debate in the 3d magnetism.$%
^{1-6}$ A non-magnetic state observed in LaCoO$_3$, for instance, is very
intriguing having in mind the strong magnetic Co state realized in CoO that
is antiferromagnet with T$_N$ of 289 K. An argument that the cobalt ion in
the high oxidation state (Co$^{3+}$) likes to be in the low-spin state does
not hold as SrCoO$_{2.5}$, where also the trivalent cobalt ion exists, shows
extremely strong antiferromagnetic state (T$_N$ of 570 K). Following Van
Vleck the realization of the low- or high-spin state is often discussed in
the one-electron model as resulting from the delicate interplay of the
crystal-field and intra-atomic exchange (Hund coupling) energies (see e.g.
Ref. 5). Despite of the numerous very different versions of the one-electron
description (LDA, LSDA, .... +GGA, ... ) this approach still has serious
difficulty in the satisfactory systematic description of e.g. the LaMO$_3$
(M=Ti-Cu) series. It produces e.g. often wrongly the metallic state instead
of the insulating/semiconducting state.$^2$

The aim of this Letter is to point out that it is the local symmetry that
determines the realization of the low- or high-spin state of the Co$^{3+}$
ion. We can prove it for the Co$^{3+}$ ion placed at the slightly distorted
octahedral site. It turns out that in the CoO$_6$ octahedra the Co$^{3+}$
ion at the absolute zero temperature can have the magnetic moment as large
as 3.6 $\mu _B$ or null as the function of the sign of the local off-cubic
distortion. In case of the rhombohedral distortion the change of the sign is
associated with the compression or the elongation of the octaedron along the
main cube diagonal. The same strong dependence of the atomic magnetic moment
holds for the tetragonal distortion.

We treat the Co$^{3+}$ ion in a solid as the highly-correlated electron
system 3d$^6$ with 6 electrons in the unfilled 3d shell. These high
correlations assure the realization of Hund's rules. We have calculated the
energy spectrum of such the system for the ground state described by the
Hund's rules quantum numbers S=2 and L=2 taking into account the spin-orbit
(s-o) coupling. We have taken the spin-orbit coupling rigorously into
account, not by approximate perturbation methods as is usually made in the
current literature, if the s-o coupling is considered at all. This 25-level
discrete energy spectrum depends on the detailed shape of the electric-field
potential formed by local charge surroundings. This detailed shape of the
electric potential can be represented by means of the multipolar expansion
of spherical harmonics. These multipolar charge interactions in the CoO$_6$
octahedra can be accounted for by consideration of the crystal-field
Hamiltonian$^7$

$H_{CF}=$B$_4^d(O_4^0-$20$\sqrt{\text{2}}O_4^3)+$B$_2^0O_2^0.$

The first term accounts for the multipolar charge interactions of the cubic
symmetry with z axis taken along the main diagonal. The second term accounts
for the simplest off-cubic distortion and $O_2^0$=3$L_z^2$-L(L+1). The cubic
term in combination with the s-o coupling, $\lambda _{s-o}$ $L\cdot S,$
yields$^8$ 3-fold degenerated ground state in the $\left| \text{LSL}_{\text{z%
}}\text{S}_{\text{z}}\right\rangle $ space with the corresponding magnetic
moments of 0 (singlet) and $\pm $3.5 $\mu _B$ (doublet). The off-cubic
distortion splits these states making the ground state magnetic (doublet) or
non-magnetic (the singlet) as the function of the sign of the B$_2^0$
parameter. The change of the sign of B$_2^0$ can be realized by the
compression or the elongation of the octaedron along the main cube diagonal
in the case of the rhombohedral distortion. The magnetic doublet forms the
long-range magnetic state in contrast to the singlet state that yields the
diamagnetic behaviour like it is in LaCoO$_3$, for instance. These
calculations have been performed with the realistic values: B$_4^d$= -11.5
meV and the spin-orbit coupling $\lambda _{s-o}$ of -18 meV. B$_2^0$%
%TCIMACRO{\TEXTsymbol{>}}
%BeginExpansion
\mbox{$>$}
%EndExpansion
0 yields the non-magnetic ground state [9]. B$_2^0$ of e.g. 2 meV produces
the spin-like gap of 1.9 meV with the highly-magnetic excited doublet with
the moment of 3.0 $\mu _B$. Such the electronic structure has been recently
suggested to exist in LaCoO$_3$ by Zhuang et al.$^6$ (m=2.9 $\mu _B$).

In conclusion, we argue that the formation of the non-magnetic or the
magnetic state of the Co$^{3+}$-ion containing compounds, discussed in the
current literature as the Co$^{3+}$ low- and high-spin states, results from
the sign of the local off-cubic crystalline-electric-field distortion. It
can be exactly calculated treating the Co$^{3+}$ ion as the
highly-correlated 3d$^6$ system that experiences the cubic crystal-field
interactions provided the spin-orbit coupling is correctly taken into
account. The proposed mechanism, the strong correlation of the local
magnetic moment and the detailed shape of the crystal-field potential
experienced by the paramagnetic ion, is known to work well for rare-earth
ion compounds [10-12] but has not been used so far for LaCoO$_3.$ We would
like to notice that the obtained correlation of the local magnetic moment
and the local symmetry is very general. In particular, it does not depend on
the used parameters provided the sign of three parameters B$_4^d$, B$_2^0$
and the spin-orbit coupling $\lambda _{s-o}$ is preserved. Moreover, the
proposed single-ion-like crystal-field-based model yields in the natural way
the insulating state experimentally-observed for most of 3d-ion compounds
like in LaCoO$_3$.

*This paper has been submitted 29.03.1999 to Phys.Rev.Lett. [LC7763] but has
been rejected by the Managing Editor (Dr G.Wells) and the\ Editor in Chief
(Dr M.Blume) in the course of a special discriminating policy with respect
to our papers. Such the policy is the manipulation of Science by the Editors
of Phys.Rev.Lett. and violates the fundamental scientific rules. Our request
to publish our paper with negative referee reports have been ignored.

\end{document}